\documentclass[pra,twocolumn,amsmath]{revtex4}
\usepackage{graphicx}

\begin{document}

\def\ket#1{|#1\rangle} 
\def\bra#1{\langle#1|}
\def\av#1{\langle#1\rangle}
\def\myarrow{\mathop{\longrightarrow}}

\title{Minimum energy needed to perform a quantum logical gate}

\author{Julio Gea-Banacloche}
\email[]{jgeabana@uark.edu}
\affiliation{Department of Physics, University of Arkansas, Fayetteville, AR 72701}

\date{\today}

\begin{abstract}
A lower bound on the amount of energy needed to carry out an elementary logical operation on a qubit system, with a given accuracy and in a given time, has been recently postulated.  This paper is an attempt to formalize this bound and explore the conditions under which it may be expected to hold.  This is a work in progress and any contributions will be appreciated.
\end{abstract}
\maketitle

\section{Introduction}

It has become of interest lately to explore the constraints that the quantum nature of the control degrees of freedom might impose on the practical operation of quantum logical gates \cite{vanenk,wang,me1,ozawa}.  A very general result derived recently by Ozawa \cite{ozawa} is that any quantum gate that changes the energy or angular momentum state of a qubit will require a minumum number of ancillary bosons of the order of $1/\epsilon$, if it is to have a failure probability smaller than $\epsilon$.  If the bosons are excitations of a quantum harmonic oscillator (such as, e.g., photons) of frequency $\omega$, this becomes a minimum energy requirement 
\begin{equation}
E_{min} \sim \frac{\hbar\omega}{\epsilon}
\label{one}
\end{equation}
in agreement with previous studies \cite{vanenk,me1} which focused on the effect of the quantum nature of the electromagnetic field on the performance of logical gates.

Ozawa's result has very wide applicability, but it must be kept in mind that it is relatively straightforward (and it may be, in fact, advantageous for other practical reasons) to encode a logical qubit in degenerate states of systems of a few qubits, which are mutually interconvertible without any energy or angular momentum cost:  for instance, the encoding in a 3-qubit decoherence-free subsystem \cite{dfs} uses as the logical zero the state $\ket 0_L = 2^{-1/2}(\ket{01} - \ket{10})\ket 0$ of three physical qubits, and as the logical one the state $\ket 1_L = 6^{-1/2}(\ket{100} + \ket{010} - 2 \ket{001})$.  These two states have the same quantum numbers for total angular momentum and energy; in fact, they simply represent the two different ways to get a state with $l=1/2$ and $m=-1/2$ in a system of three spin-$1/2$ particles.  For such an encoding, conservation of total energy or angular momentum alone does not appear to restrict the possible logical operations. 

I have recently shown \cite{me2} that in many cases, regardless of whether a conservation law is broken or not by the action of the logical gate, there is a minimum requirement on the energy of the ``control'' system, or degree of freedom, of the form (1) if the system is an oscillator, or more generally of the form
\begin{equation}
E_{min} \sim \frac{\hbar}{\epsilon T}
\label{two}
\end{equation}
if the gate is to be carried out in a time $T$ with failure probability less than $\epsilon$.  My analysis covers gates mediated by external electromagnetic fields, or by controlled collisions between particles, assuming that the fields or particles are in minimum uncertainty ``coherent states.'' There are, nonetheless, some questions still open, regarding the full generality of the result, and, for instance, whether placing the control degree of freedom in a nonclassical state (such as a squeezed state) might lower the bounds or not.  In this paper, which is offered to the community as a working document, I shall attempt to express the constraint (2) as a formal postulate, and exhibit a number of worked out examples and ideas for how a general proof might proceed.  Any help or suggestions will be greatly appreciated.  

The reader with only a casual interest may start by skipping the detailed calculations of Sections 5 and 6 and reading only the more heuristic arguments in the other sections.  

\section{A formal postulate}

In order to focus only on the constraints arising from the quantum nature of the control, and not on those imposed by conservation laws, I have focused on a particular kind of two-qubit gate which preserves the qubits' energy and angular momentum (assuming the $\ket 0$ and $\ket 1$ are eigenstates of these variables), namely, the controlled sign-flip gate, which leaves the states $\ket{00}, \ket{01}$ and $\ket{10}$ unchanged but turns $\ket{11}$ into $-\ket{11}$.  The role of the control system is, essentially, to switch ``on'' and ``off'' a Hamiltonian which accomplishes this in a time $T$, to an accuracy given by $\epsilon$.  

To that end, let the control degree of freedom be initially in the state $\ket{\psi_0}$, and let its self-Hamiltonian be $H_0$.  Let the interaction Hamiltonian have the simple form $H_I = V \ket{11}\bra{11}$.  This is the minimal form needed for the purpose at hand (question:  would it be worth it to consider more complicated couplings?) and $V$ need depend only on ``control'' operators.  Further, suppose $V$ is time-independent in the Schr\"odinger picture, although this may not be necessary.  What {\it is\/} necessary is that the interaction be turned on and off only by the control system, acting under the influence of its own self-Hamiltonian (it follows that $\ket{\psi_0}$ cannot be a stationary state).  Formally, we require
\begin{equation}
\av{\psi_0 | V^2| \psi_0} \simeq \bra{\psi_0} e^{\frac{i}{\hbar} \int_0^T  H_0 \, dt'}| V^2| e^{-\frac{i}{\hbar} \int_0^T  H_0 \, dt'} \ket{\psi_0}\simeq 0
\label{three}
\end{equation}
at the initial and final times, $t=0, T$.  
   
To capture the desired change in sign, we define the ``failure probability'' of the gate by considering what it does to a state such as $\ket{00}+\ket{11}$.  What we want is something like
\begin{widetext}
\begin{equation}
\frac{1}{\sqrt 2} (\ket{00}+\ket{11})\ket{\psi_0} \to \frac{1}{\sqrt 2} (\ket{00}-\ket{11})e^{-\frac{i}{\hbar} \int_0^T  H_0 \, dt'}\ket{\psi_0}
\label{four}
\end{equation}
What we will get is, instead,
\begin{equation}
\frac{1}{\sqrt 2} (\ket{00}+\ket{11})\ket{\psi_0} \to \frac{1}{\sqrt 2} \ket{00}e^{-\frac{i}{\hbar} \int_0^T  H_0 \, dt'}\ket{\psi_0} + \frac{1}{\sqrt 2}\ket{11} e^{-\frac{i}{\hbar} \int_0^T  (H_0 + V) \, dt'}\ket{\psi_0}
\label{five}
\end{equation}
\end{widetext}
and the ``failure probability'' can therefore be defined as 1 minus the square of the overlap between (4) and (5), i.e.,
\begin{eqnarray}
p &&= 1 -  \frac{1}{4} \left|1 -  \bra{\psi_0} e^{\frac{i}{\hbar} \int_0^T  H_0 \, dt'} e^{-\frac{i}{\hbar} \int_0^T  (H_0 + V) \, dt'}\ket{\psi_0} \right|^2 \nonumber \\
&&= 1 -  \frac{1}{4} \left|1 - \bra{\psi_0} {\cal T} e^{-\frac{i}{\hbar} \int_0^T  V_I(t') \, dt'}\ket{\psi_0} \right|^2
\label{six}
\end{eqnarray}
where the last equation is written in the interaction picture, and time-ordering is denoted by ${\cal T}$.
Now we have all the ingredients needed to make a general, formal claim:  in order to be able to turn on and off an interaction strong enough to flip the sign of the wavefunction in (6) over the time $T$, and to do this accurately enough, so that $p < \epsilon$ (where $\epsilon$ is some acceptable error) the state $\ket{\psi_0}$ must have a minimum energy of the order of 
\begin{equation}
\av{\psi_0 |H_0 | \psi_0}_{min}  \sim \frac{\hbar}{\epsilon T}
\label{seven}
\end{equation}
This claim involves only the (arbitrary) control system, its self-Hamiltonian, and the interaction $V$.  The question is, how generally can this be established?

\section{A counterexample to show that the condition (3) is necessary}

At the risk of belaboring the obvious, here is a simple example showing that the constraint disappears if one allows the interaction to be always ``on,'' that is, if (3) does not hold.  Let the control degree of freedom be a harmonic oscillator, let $\ket{\psi_0} = \ket{n}$, an energy eigenstate, and let $V= \hbar g a^\dagger a$.  Then one only has to choose $T=\pi/gn$ and equation (6) will be satisfied exactly, with $p=0$, which means one could make $\epsilon$ arbitrarily small, and (7) would be violated.

\section{Some insights on the reason for the constraint}

At this point it is natural to ask, why, then, is it impossible to make $p$ in (6) exactly zero if eqs. (3) are enforced?  A first answer is that, clearly, if $V_I(t)\ket{\psi_0}$ was nearly equal to zero at all times the interaction would have essentially no effect; so then $\ket{\psi_0}$ cannot be an eigenstate of $V_I(t)$, which means that $V_I(t)$ will not be sharply defined in the state $\ket{\psi_0}$.  There will be fluctuations, which one could formally separate out as 
\begin{equation}
V_I(t) = \av{V_I(t)} + \Delta V_I(t)
\label{eight}
\end{equation}
with $\av{V_I(t)} \equiv \bra{\psi_0} V_I(t) \ket{\psi_0}$, and $\ket{\psi_0}$ not an eigenstate of $\Delta V$.  In that case, if one chooses $V$, $\ket{\psi_0}$, and $T$ so that
\begin{equation}
\int_0^T \av{V_I(t')} dt' = \pi\hbar
\label{nine}
\end{equation}
one may estimate the failure probability $p$ given by (\ref{six}), by expanding the exponential, as 
\begin{equation}
p\simeq \frac{1}{2}\int_0^T dt \int_0^T dt' \bra{\psi_0} \Delta V_I(t) \Delta V_I(t') \ket{\psi_0}
\label{ten}
\end{equation}
and this is the approach that was adopted in \cite{me2} and in most of the examples to follow.  (The time ordering has been dropped in reaching (\ref{ten}), which is presumably not important if all one wants is an estimate.)

On the other hand, perhaps a more fundamental way to look at this is to realize that the presence of the operators $\Delta V$ in the exponent of (\ref{six}) {\it changes\/} the state $\ket{\psi_0}$, so that it no longer exactly overlaps with itself, and (\ref{ten}) simply estimates the extent of this mismatch.  From this point of view, what we have here is just the old idea that, in order to be able to observe interference in a quantum mechanical system that is interacting with a classical ``apparatus,'' the apparatus (in this case, the ``control'' system described by $\ket{\psi_0}$) must be large enough for the ``back reaction'' of the quantum system on it to be negligible.

One way to think along these lines may be as follows.  Putting together (\ref{three}) and (\ref{six}), one can say that the action of the self-Hamiltonian $H_0$ on $\ket{\psi_0}$ changes it, in a time of the order of $T$, from a state for which $V\ket{\psi_0} \simeq 0$ to a state for which $V\ket{\psi_0}$ is of the order of $(\pi\hbar/T)\ket{\psi_0}$ (this is a measure of the degree of noncommutativity between $H_0$ and $V$).  One may expect the $V$ in the exponent of (\ref{six}) to have a similar effect, then, and ``displace'' the energies of the states making up $\ket{\psi_0}$ by an approximate amount $\Delta E \sim \pi\hbar/T$.  The condition (\ref{seven}), then, would express the minimum energy that $\ket{\psi_0}$ must have in order to still overlap with itself to the degree given by $\epsilon$, after its component energies have been ``messed up'' by an amount of the order of $\Delta E$; in this language, it simply reads $\Delta E/E \le \epsilon$. 

The next two sections contain a number of worked out examples, essentially applying Eq.~(\ref{ten}) to various situations and showing how something like (\ref{seven}) arises in every case.  After that, the last Section reexamines the case of material particles (where the ``control'' degree of freedom is, say, a particle's center-of-mass coordinate) in a heuristic manner, and presents some final thoughts.      
 
\section{Switching by means of a (quantized) e.m. field}

\subsection{Linear coupling}
Assume linear coupling, and a multimode coherent state:
\begin{equation}
H_{I} =\hbar \left( \sum_{k}g_{k}a_{k}e^{-i\omega_{k}t} + H.c \right)\ket{11}\bra{11}
\end{equation}
\begin{equation}
\ket{\psi}=\prod_{k}\ket{\alpha_{k}}
\end{equation}
with
\begin{equation}
\int \sum_{k}g_{k}\alpha_{k}e^{-i\omega_{k}t} dt +c.c. = \pi 
\label{eleven}
\end{equation}

Error arises from quantum fluctuations in the amplitude of the field modes.  To keep it smaller than $\epsilon$ one will require
\begin{equation}
\sum_{k}\left| \int g_{k}e^{-i\omega_{k}t}dt \right|^{2} < \epsilon \label{twelve}
\end{equation}
However, from (\ref{eleven}) we get that
\begin{equation}
\sum_k |\alpha_{k}| \left|\int g_{k}e^{-i\omega_{k}t}dt \right| \ge \frac{\pi}{2}
\end{equation}
Then (\ref{twelve}) is only possible if
\begin{equation}
\left(\sum_k |\alpha_{k}|^{2} \right)^{1/2} \ge \frac{\pi}{ 2\sqrt{\epsilon}}
\end{equation}

The pulse's ``average frequency'' is
\begin{equation}
\av{\omega} = \frac{\sum_k \omega_k |\alpha_k|^2}{\sum_k |\alpha_k|^2} \le \frac{4 \epsilon}{\pi^2 } \sum_k \omega_k |\alpha_k|^2
\end{equation}
and the total pulse energy is
\begin{equation}
E_{field} = \sum_k \hbar \omega_k |\alpha_k|^2
\end{equation}
so
\begin{equation}
E_{field}\ge \frac{\pi^2}{4}\, \frac{\hbar\av{\omega}}{\epsilon} 
\label{thirteen}
\end{equation}
Note that for a ``static'' field, switched on and off over a time $T$, $\av{\omega} \sim 1/T$

\subsection{Nonlinear coupling}
Assume the Hamiltonian is of the form:
\begin{equation}
H_{I} =\hbar g E^p(t) \ket{11}\bra{11}
\end{equation}
with $g$ a time-independent coupling constant.  Let 
\begin{equation}
E(t) = \sum_k \sqrt{\frac{\hbar\omega_k}{2 \epsilon_0 V}} a_k e^{-i\omega_k t} + H.c
\end{equation}
It is understood that the sum over frequencies is limited by the natural frequency response of the system.  Let ${\cal E}(t) = \bra\psi E \ket\psi$ and $E = {\cal E} + \Delta E$.  Then
\begin{eqnarray}
\Delta H &&= \hbar g p {\cal E}^{p-1} \Delta E \nonumber \\
&&= \hbar g p {\cal E}^{p-1} \nonumber \\
&&\quad\times\sum_k \sqrt{\frac{\hbar\omega_k}{2 \epsilon_0 V}} \left(\Delta a_k e^{-i\omega_k t} + \Delta a_k^\dagger e^{i\omega_k t} \right)
\end{eqnarray}
The error in the operation of the gate can be estimated as
\begin{equation}
\left\langle \left(\int \Delta H dt/\hbar \right)^2 \right\rangle = \sum_k \left| p g \sqrt{\frac{\hbar\omega_k}{2 \epsilon_0 V}} \int {\cal E}^{p-1} e^{-i\omega_k t} dt \right|^2
\end{equation} 
whereas, on the other hand, we want
\begin{equation}
g \int {\cal E}^p dt = \pi
\end{equation}
and the left-hand side of this expression can be written as
\begin{equation}
\sum_k g \sqrt{\frac{\hbar\omega_k}{2 \epsilon_0 V}} \int {\cal E}^{p-1} \alpha_k e^{-i\omega_k t} dt + c.c.
\end{equation}
The two conditions
\begin{equation}
\sum_k \alpha_k  \int  g\sqrt{\frac{\hbar\omega_k}{2 \epsilon_0 V}} {\cal E}^{p-1} e^{-i\omega_k t} dt + c.c. = \pi
\end{equation}
and 
\begin{equation}
\sum_k \left| \int g \sqrt{\frac{\hbar\omega_k}{2 \epsilon_0 V}} {\cal E}^{p-1} e^{-i\omega_k t} dt \right|^2 < \frac{\epsilon}{p^2}
\end{equation}
are formally equivalent to (\ref{eleven}) and (\ref{twelve}) with a time-dependent $g_k$ and a modified $\epsilon$, and so the same logic applies to yield for the total field energy
\begin{equation}
E_{field}\ge \frac{\pi^2}{4}\, \frac{p^2 \hbar\av{\omega}}{\epsilon}
\end{equation} 
As long as $p^2 \ge 1$ (i.e., the Hamiltonian is an analytic function of the field) this condition is at least as restrictive as (\ref{thirteen}).

\subsection{Squeezing?}

In a coherent state, both ``quadratures'' of the field-amplitude operator $a_k$ have the same noise.  One could imagine a Hamiltonian that couples only to one quadrature, which could then be squeezed.  

What might happen then could be roughly as follows.  The fluctuations (squared) in the squeezed quadrature would be reduced by a factor $e^{-2r}$, where $r$ is the squeezing parameter.  This could amount to formally increasing $\epsilon$ in Eq.~(\ref{twelve}) by a factor $e^{2r}$.  Note that the number of photons in the field is now given by $|\alpha_k|^2 + e^{2r}$, so, in fact, the equation for the field energy might end up reading
\begin{equation}
E_{field} \ge {\hbar \av{\omega}}\left(\frac{1}{e^{2r} \epsilon} +  e^{2r} \right)
\end{equation}
When this expression is minimized over $r$, one obtains
\begin{equation}
E_{field} \ge \frac{2 \hbar \av{\omega}}{\sqrt{\epsilon}} 
\label{oldfour}
\end{equation}
Note, however, that to couple to a squeezed field one typically needs a local oscillator at the carrier frequency $\omega$.  Presumably, if $\omega$ is not sufficiently sharply defined, errors in the gate operation will result.  This means that broadening of $\omega$ due to the finite pulse duration must be prevented.  If the condition
\begin{equation}
(\omega T)^2 > \frac{1}{\epsilon}
\end{equation}
is applied to equation (\ref{oldfour}), one obtains again
\begin{equation}
E_{field} \ge \frac{\hbar}{\epsilon T} 
\label{oldfive}
\end{equation}
Thus, it seems that even using squeezing one is still constrained by the inequality (\ref{oldfive}).  A more careful study of this possibility, however, may be necessary, ideally in the context of a specific model for the coupling interaction.

\section{Switching using collisions between wavepackets}

\subsection{``Free'' particles}
Suppose one arranges to have a collision between the two particles involved in the gate operation, with the idea that their mutual interaction, $V(|{\bf r}_1 - {\bf r}_2|)$, will provide the desired phase shift.  Work in the center of mass frame assuming identical particles; neglect deviations of the particles' motion from straight lines at constant speed; let $b$ be the distance of closest approach and take that to be the $x$ direction.  Then what we want is  
\begin{equation}
 \frac{1}{\hbar} \int_{-T/2}^{T/2} V(\sqrt{4v^2 t^2 + b^2}) dt = \pi \label{oldsix}
\end{equation} 
where $\pm v$ is the $y$-component of the particles' velocity in the CM frame.  

In practice the free wavepackets' $x$ coordinate is uncertain by an amount equal to
\begin{equation}
\Delta x(t) = \Delta x_0 + \frac{\Delta p_0}{m} \, \left(t + \frac{T}{ 2}\right)
\label{oldseven}
\end{equation} 
assuming that $x$ and $p$ are initially uncorrelated (at the time $t=-T/2$).  This alone causes an uncertainty in (\ref{oldsix}) whose average magnitude square goes as
\begin{equation}
\delta^2  = \frac{b^2}{\hbar ^2} \left(\int \frac{dV}{d\rho} \, \frac{dt}{ \rho} \right)^2  \left(\Delta x_0^2 +  \frac{T^2 \Delta p_0^2}{4 m^2} \right)
\end{equation} 
with $\rho = (4 v^2 t^2 + b^2)^{1/2}$, and making use of the symmetry of the integrands.  

One can now minimize $\delta^2$ with the constraint $\Delta x_0 \Delta p_0 \ge \hbar/2$ (i.e., pick an optimal wavepacket), with the result $\Delta x_0^2 = T\hbar/4 m$, $\Delta p_0^2 = 2 m \hbar/T$.  Then the condition $\delta^2 <\epsilon$ becomes
\begin{equation}
\frac{b^2}{\hbar ^2} \, \frac{T\hbar}{2m} \left(\int \frac{dV}{d\rho} \, \frac{dt}{ \rho} \right)^2  <\epsilon
\end{equation} 
Now use (\ref{oldsix}) to eliminate the first factor of $1/\hbar^2$, and consider the derivative of the left-hand side of (\ref{oldsix}) with respect to the impact parameter $b$.  One easily obtains the constraint
\begin{equation}
\frac{\pi^2 T\hbar}{2m} \left[\frac{d}{d b} \ln \left(\int_{-T/2}^{T/2} V(\sqrt{4v^2 t^2 + b^2}) dt \right) \right]^2 < \epsilon 
\end{equation} 
This can be simplified a bit by, first, introducing the obvious change of variable $2vt = y$, with limits of integration $\pm y_0 = \pm v T$, and then assuming that $y_0$ is large enough that no substantial error is introduced by extending the integration to $\pm \infty$.  This is reasonable, since the particles have to start and end far enough away from each other for their mutual interaction to be negligible.  We then have
\begin{equation}
\frac{\pi^2 T\hbar}{2m} \left[\frac{d}{d b}  \ln \left(\int_{-\infty}^{\infty} V(\sqrt{y^2 + b^2}) dy \right) \right]^2 < \epsilon  \label{oldeight}
\end{equation}
The left-hand side of (\ref{oldeight}) is easily evaluated when $V(\rho)$ is any power law, since one can write $(y^2+b^2)^{-n/2} = b^{-n} ((y/b)^2+1)^{-n/2}$ and then the change of variable $y/b = u$ results in a factor of $b^{-n+1}$ times an integral which is independent of $b$.  Hence for any $n>1$, we get
\begin{equation}
\frac{\pi^2 T\hbar}{2m}\, \frac{(n-1)^2}{ b^2} < \epsilon
\end{equation}
and, since, as argued above, we must have $b < y_0 = v T$, this yields immediately
\begin{equation}
\frac{\hbar}{ m v^2 T} < \epsilon
\end{equation}
or
\begin{equation}
mv^2 > \frac{\hbar}{\epsilon T}
\end{equation}
where $mv^2$ is the initial kinetic energy of the two particles.  

\subsection{Particles in a harmonic potential}

To prevent spreading of the wavepackets during the interaction, one could imagine confining the particles in a static potential (a time-dependent potential implies a time-dependent field, and we are back to the previous Section).  Assume the potential is harmonic, and consider the following scenario:  at time $t=0$ we create the two wavepackets, a distance $4A+b$ apart, let them oscillate towards each other with amplitudes $A$, so that, at the time $t=\pi/\omega$, they are closest, a distance $b$ apart; then they swing back to the starting position by the time $t= 2\pi/\omega$.  At a minimum, one needs to put in and remove enough energy to start and stop this pendulum motion.  Just how this is done is left vague for the moment, but it might be important later on.

In any case, assume that we have an interaction energy $V(\rho)$, as before, but now only in one dimension, with
\begin{eqnarray}
x_1 &&= -\left(A+\frac{b}{2}\right) - A \cos\omega t \nonumber \\
x_2 &&= \left(A+\frac{b}{2}\right) + A \cos\omega t 
\end{eqnarray}
and $\rho = x_2-x_1 = 2A + b + 2A \cos\omega t$.  
The desired action is
\begin{equation}
\frac{1}{\hbar} \int_0^{2\pi/\omega} V(\rho) dt = \pi
\end{equation}
The error operator $\Delta x$ for a harmonic oscillator is
\begin{equation}
\Delta x = \Delta x_0 \cos\omega t + \Delta p \sin\omega t
\end{equation}
so we find, proceeding as before,
\begin{widetext}
\begin{eqnarray}
\delta^2 &&= \frac{2}{\hbar^2} \left[\left(\int \frac{dV}{d\rho} \cos \omega t \,dt\right)^2 \Delta^2 x_0 + \left(\int \frac{dV}{d\rho} \sin \omega t \,dt\right)^2 \Delta^2 p_0 \right] \nonumber \\
&& = \frac{2}{\hbar^2} \left(\int \frac{dV}{d\rho} \cos \omega t \,dt\right)^2 \Delta^2 x_0 
\label{fortyone}
\end{eqnarray}
\end{widetext}
The contribution of $\Delta p_0$ vanishes due to the symmetry of the integration, which raises again the possibility of using squeezing to improve on the constraint to be derived presently.  For the moment, however, assume simply that a coherent state wavepacket is excited; then $\Delta x_0$ is just the ordinary zero-point fluctuation of the ground state of a harmonic oscillator, $\Delta^2 x_0 = \hbar/2m\omega$, and now the constraint we have is
\begin{equation}
\frac{\pi^2 \hbar}{ m\omega} \frac{\displaystyle \left(\int_0^{2\pi/\omega}\frac{dV}{ d\rho} \cos \omega t \,dt\right)^2} {\displaystyle \left(\int_0^{2\pi/\omega} V(\rho) \,dt \right)^2} < \epsilon
\end{equation}
The integrals in this case do not seem so easy to evaluate in general, but specific cases can readily be done.  For instance, for a dipole-dipole $\rho^{-3}$ interaction one finds to leading order in $b$
\begin{equation}
\frac{\pi^2 \hbar}{ m\omega}  \left(\frac{5}{2 b} \right)^2 < \epsilon
\end{equation}
Note that the energy of each oscillator is $\frac{1}{2} m\omega^2 A^2$, that the interaction time $T = 2\pi/\omega$, and that, as before, we'll want $A > b$, so again we find that
\begin{equation}
m\omega^2 A^2 > \frac{\hbar}{\epsilon T} 
\label{oldnine}
\end{equation}  

As mentioned above, it looks as if one could use a squeezed state (squeezed in the position variable) to improve on the constraint (\ref{oldnine}).  This is because $\Delta^2p_0$ does not appear in Eq.~(\ref{fortyone}), which in turn follows from the symmetry of the integral over the {\it unperturbed\/} trajectory.  Numerical calculations done for a classical particle, however, show that (as is only to be expected), as a result of the interaction, the particle does not return exactly to the starting point.  The importance of this mismatch between the perturbed and unperturbed wavepackets would only be magnified if the quantum wavepacket was squeezed in position.  Hence, there has to be a limitation to how much one can squeeze the position, but it does not seem a simple matter to derive it.  Specifically, it seems that, in  the formalism used here, these effects would appear to a higher order in the expansion (\ref{fortyone}).  

On the other hand, there may be good (self-consistency) reasons to go to higher orders in (\ref{fortyone}) in the case of large squeezing.  If one has a state that is squeezed in position, say, enough to change the dependence of the minimum energy on $\epsilon$, from $\epsilon^{-1}$ to $\epsilon^{-1/2}$ (the best achievable in any case, by the arguments of Section V.C),  the squeezing factor $e^{-2r}$ in $\Delta^2x_0$ would have to be of the order of $\sqrt{\epsilon}$.  In that case, the corresponding factor $e^{2r}$ in $\Delta^2p_0$ would be of the order of $1/\sqrt{\epsilon}$, and in $\Delta^4p_0$ it would be of the order of $1/\epsilon$.  This suggests that in case of such extreme squeezing, one would not be justified to neglect the higher order terms (in particular, terms of order $\Delta^4p_0$) in (\ref{fortyone}). I am currently looking into this.

\section{Discussion}

For material particles, there is actually a very easy way to ``derive'' the constraint (\ref{two}) heuristically, based on some of the ideas presented in Section 4.  The potential $V$ produces a force $dV/dx$ on the particle, which, acting over a time $T$, results in a position change (relative to the unperturbed wavepacket) of
\begin{equation}
\delta x \sim \frac{1}{2m}\frac{dv}{dx} T^2
\end{equation}
and a momentum change
\begin{equation}
\delta p \sim \frac{dv}{dx} T
\end{equation}
From the condition (\ref{nine}) one can estimate $V$ as $\pi\hbar/T$ and $dV/dx$ as $\pi\hbar/LT$, where $L$ is a characteristic length, that the particle traverses in the time $T$ (so the velocity $v\sim L/T$).  Presumably, the position and momentum change will lead to a ``mis-overlap'' with the original wavepacket of the order of $(\delta x/\Delta x)^2$ and $(\delta p/\Delta p)^2$, where $\Delta x$ and $\Delta p$ are the original, intrinsic position and momentum uncertainty.  The constraint
\begin{equation}
\left(\frac{\delta x}{\Delta x}\right)^2 + \left(\frac{\delta p}{\Delta p}\right)^2 < \epsilon
\end{equation}
then becomes
\begin{equation}
\left(\frac{\pi\hbar}{L} \right)^2 \left(\frac{T^2}{4 m^2} \, \frac{1}{\Delta^2 x} + \frac{1}{\Delta^2 p} \right) < \epsilon
\label{fortyeight}
\end{equation}
Using the fact that $\Delta x\Delta p \ge \hbar/2$ to optimize (minimize) the left-hand side of (\ref{fortyeight}), we find that it reduces to
\begin{equation}
\frac{2\pi^2 \hbar T}{m L^2} < \epsilon
\end{equation}
which is to say
\begin{equation}
\frac{1}{2}mv^2 > \frac{\pi^2\hbar}{T}
\end{equation}
if $v \sim L/T$.  

The simplicity of this argument is in sharp contrast with the complexity of the specific examples worked out in the previous section.  One must, therefore, ask, is there a formally simple way to make the above heuristic argument rigorous and general?  

It is probably not hard to see in the case of the electromagnetic field (Section 4) a generalization of the above argument to a situation dealing with many ``generalized coordinates and momenta.'' One may also ask, at this point, whether it is essential, for an inequality of the form (2) to hold, that the control system's self-energy be a quadratic function of the generalized coordinates and momenta.

Finally, in the case of an electromagnetic pulse the pulse duration comes out naturally from the formalism of Section 5, whereas for the material particles of Section 6 there is some ambiguity as to what to use for $T$.  It might be nice (necessary?) to define $T$ formally in some way, perhaps using the Fourier transform of $\av{V_I(t)}$.

\end{document}